\documentclass[aps,prd,preprint,groupedaddress,longbibliography]{revtex4-1}

\usepackage{amsmath,amssymb,amsfonts}
\usepackage[T1]{fontenc}
\usepackage{graphicx}
\usepackage{array}
\newcolumntype{L}{>{\centering\arraybackslash}p{3cm}}
\usepackage{hyperref}

\begin{document}

\title{Dark energy and spontaneous mirror symmetry breaking}


\author{Wanpeng Tan}
\email[]{wtan@nd.edu}
\affiliation{Department of Physics, Institute for Structure and Nuclear Astrophysics (ISNAP), and Joint Institute for Nuclear Astrophysics - Center for the Evolution of Elements (JINA-CEE), University of Notre Dame, Notre Dame, Indiana 46556, USA}

\date{\today}

\begin{abstract}
Dark energy is interpreted as the leftover of mostly canceled vacuum energy due to the spontaneous mirror symmetry breaking (SMSB) at the electroweak phase transition. Based on the newly proposed mirror-matter model (M$^3$), the extended standard model with mirror matter (SM$^3$) is elaborated to provide a consistent foundation for understanding dark energy, dark matter, baryogenesis, and many other puzzles. New insights of Higgs, top quark, and lepton masses are presented under SM$^3$ using staged quark condensation and four-fermion interactions for SMSB. In particular, the nature and mass scales of neutrinos are naturally explained under the new theory. The new cosmology model based on SM$^3$ could potentially resolve more cosmic enigmas. The possible underlying principles for SMSB and SM$^3$ of a maximally interacting, supersymmetric, and mirrored world are also discussed.
\end{abstract}


\maketitle

\section{Introduction\label{intro}}

The contents of our universe are dominated by dark energy, which was first evidenced by the discovery of an accelerating universe with type Ia supernovae \cite{riess1998,perlmutter1999}. Many other probes such as cosmic microwave background (CMB) and baryon acoustic oscillations have more firmly supported its existence ever since and current observational evidence for dark energy can be found in recent reviews \cite{weinberg2013,huterer2018}. The standard cosmology model ($\Lambda$CDM) fitted with the current Planck2018 CMB data \cite{planckcollaboration2018} indicates that dark energy constitutes about 68\% of the total energy of the universe. A plethora of dark energy models have been proposed involving various exotic treatments like imaginary scalar fields or modified general relativity and discussed in extensive reviews \cite{copeland2006,frieman2008,li2011,bamba2012,brax2018}. However, the cosmological constant $\Lambda$ or vacuum energy \cite{weinberg1989} remains the simplest and most plausible candidate for dark energy. In particular, $\Lambda$CDM, based on the concepts of $\Lambda$ and cold dark matter, is still the best in agreement with observation \cite{wen2018}.

The Einstein field equations can generally be written as,
\begin{equation}\label{ein}
R_{\mu\nu} - \frac{1}{2}Rg_{\mu\nu} = 8\pi G T_{\mu\nu} + \Lambda g_{\mu\nu}
\end{equation}
where $\Lambda$ is the celebrated cosmological constant and the energy-momentum tensor $T_{\mu\nu}$ describes the other contents of radiation and matter (including dark matter). Using the Friedmann–Robertson–Walker (FRW) metric of $ds^2 = dt^2 - a^2(t)(dr^2/(1-kr^2)+r^2d\Omega)$ for a homogeneous and isotropic universe, we can obtain the Friedmann equation,
\begin{equation}
H^2 \equiv \left(\frac{\dot{a}}{a}\right)^2 = \frac{8\pi G}{3}(\rho_m + \rho_{rad} + \rho_\Lambda) - \frac{k}{a^2}
\end{equation}
where $H$ is the Hubble expansion rate, $a(t)$ is the FRW scale parameter, the curvature $k=0$ for a flat universe, and $\rho_m$, $\rho_{rad}$, and $\rho_\Lambda = \Lambda/8\pi G$ are densities for matter, radiation, and dark energy, respectively. The cosmological constant or dark energy has negative pressure of $p_\Lambda = -\rho_\Lambda$ and hence causes the expansion of the universe to accelerate. Current observations have constrained the dark energy density to be finite but surprisingly small, i.e., $\rho_{\Lambda} \simeq (2.3\times 10^{-3}\, \text{eV})^4$.

In quantum field theory, vacuum energy, showing the exact properties of dark energy, can be provided only by scalar fields. In the Standard Model (SM) of particle physics, the only known scalar field is the Higgs field $\phi$ which is governed by the Lagrangian
\begin{equation} \label{higgs}
\mathcal{L}_{\text{H}} = |D \phi|^2 + \frac{1}{2}m^2|\phi|^2 - \lambda \phi^4
\end{equation}
where the kinetic and mass terms describe the excitation of the field, i.e., part of the energy-momentum tensor $T_{\mu\nu}$ in Eq. (\ref{ein}). Only the last term of Eq. (\ref{higgs}) contributes to the vacuum energy. Therefore, the vacuum energy density in SM can be estimated for a scalar field as,
\begin{equation}\label{eq_rvac}
\rho_{\text{vac}} = \lambda \langle \phi \rangle^4 = \frac{\lambda}{4} v^4
\end{equation}
where $\lambda$ is a dimensionless constant and the vacuum expectation value (VEV) of the SM Higgs is $v=246$ GeV pertinent to the energy scale of the electroweak phase transition. Therefore, the dark or vacuum energy density can be simply related to the VEV of a scalar field by $\rho_\Lambda \equiv \rho_{\text{vac}} \sim v^4$ or $(\sum v_i)^4$ for multiple fields.

The SM interpretation of vacuum energy as dark energy has a serious flaw. It predicts a vacuum energy density of $\rho_{\text{vac}} \sim (10^{11}\, \text{eV})^4$ that is many orders of magnitude larger than the observed one of $\rho_{\Lambda} \sim (10^{-3}\, \text{eV})^4$. Even considering the lower quantum chromodynamics (QCD) phase transition scale of about $10^{8}$ eV, it is still too high. This is the so-called fine tuning or naturalness problem in the cosmological constant or vacuum energy interpretation. Another issue is the so-called cosmic coincidence problem, i.e., why are dark energy and matter contents of the universe on the same order now as their dynamics throughout the history of the universe is dramatically different? These puzzling questions have prompted many studies of alternative dark energy models which have their own faults \cite{copeland2006,frieman2008,li2011,bamba2012,brax2018}.

There are other even higher speculated energy scales (e.g., $\gtrsim 10^{4}$ GeV for SUSY, $10^{16}$ GeV for GUT, and $10^{19}$ GeV for the Planck scale) that can make the fine-tuning issue worse. One naive idea would be to require an almost preserved symmetry to cancel most of the contributions from scalar fields resulting in a tiny vacuum or dark energy density. This idea looks more promising in light of a recent study that has proved that global symmetries can not be perfectly conserved under the general principles of quantum gravity \cite{harlow2019}.

In this work, we shall present that the slightly broken mirror symmetry indeed provides the mechanism for the large cancellation of vacuum energy. A relative breaking scale of $10^{-15} \text{--} 10^{-14}$ for spontaneous mirror symmetry breaking (SMSB) has been proposed for consistent understanding of the neutron lifetime anomaly, dark-to-baryon matter ratio, baryogenesis, evolution of stars, ultra-high energy cosmic rays, and the extended Cabibbo-Kobayashi-Maskawa (CKM) matrix \cite{tan2019,tan2019a,tan2019b,tan2019c,tan2019d}. If SMSB occurs at the energy scale of the electroweak phase transition, a breaking scale of $10^{-14}$ could suppress the vacuum energy density exactly to $(10^{-3}\, \text{eV})^4$ in agreement with today's observation. As it turns out, dark energy and dark matter may stem from the same origin -- SMSB.

Based on the new Mirror-Matter Model (M$^3$) \cite{tan2019,tan2019a,tan2019b,tan2019c,tan2019d}, we shall first construct the extended Standard Model with Mirror Matter (SM$^3$) for consistent understanding of dark energy and other phenomena. A further view of the vacuum structure is presented with staged quark condensation and four-fermion interactions and clues about masses of leptons and the nature of neutrinos are revealed. In the end, we speculate that a maximally interacting, supersymmetric, and mirrored world may be the underlying principles for SMSB and SM$^3$.

\section{Extended Standard Model with Mirror Matter (SM$^3$) and Dark Energy\label{sm3}}

The mirror matter idea was originated from the discovery of parity violation in the weak interaction by Lee and Yang \cite{lee1956}. It has subsequently been developed into an intriguing mirror-matter theory by various efforts \cite{kobzarev1966,blinnikov1983,kolb1985,hodges1993,berezhiani2004,berezhiani2006,cui2012,foot2014}. The general picture of the theory is that a parallel sector of mirror particles exists as an exact mirrored copy of the known ordinary particles and the two worlds can only interact with each other gravitationally. Nevertheless, many of previous models \cite{berezhiani2004,berezhiani2006,cui2012,foot2014} attempted to add some explicit feeble interaction between the two sectors. On the contrary, in the newly proposed mirror-matter model (M$^3$) \cite{tan2019}, no explicit cross-sector interaction is introduced, namely, the two parallel sectors share nothing but the same gravity before the mirror symmetry is spontaneously broken. Based on M$^3$, we shall extend the Standard Model with gauge symmetry group of $G = SU(3)_c \times SU(2)_L \times U(1)_Y$ by adding the mirror counterpart $G'=SU(3)'_c \times SU(2)'_R \times U(1)'_Y$.  
As presented below, we shall clarify misconceptions of the mirror symmetry transformation and elaborate the extended Standard Model with Mirror Matter (SM$^3$) in a rather exact and consistent form.

{
The Standard Model (SM) Lagrangian can be summed from the contributions of $SU(3)_c$ QCD, electroweak (EW) with gauge group of $SU(2)_L\times U(1)_Y$, and Higgs as follows,
\allowdisplaybreaks
\begin{eqnarray}
\mathcal{L}_{\text{SM}} &=& \mathcal{L}_{\text{QCD}} + \mathcal{L}_{\text{EW}} + \mathcal{L}_{\text{Higgs}},\\
\mathcal{L}_{\text{QCD}} &=& -\frac{1}{4}G^a_{\mu\nu}G^{a\mu\nu} + \sum^6_{j=1} (\bar{q}^L_j i\gamma^{\mu}D_{\mu} q^L_j + \bar{q}^R_j i\gamma^{\mu}D_{\mu} q^R_j),\\
\mathcal{L}_{\text{EW}} &=& -\frac{1}{4}W^b_{\mu\nu}W^{b\mu\nu} - \frac{1}{4}B_{\mu\nu} B^{\mu\nu} + \sum_{j} ( \bar{L}_j i\gamma^{\mu}D^L_{\mu} L_j + \bar{R}_j i\gamma^{\mu}D^R_{\mu} R_j ),\\
\mathcal{L}_{\text{Higgs}} &=& -\sum_j y_j(\bar{\psi}^L_j \psi^R_j\phi + \bar{\psi}^R_j \psi^L_j \phi^\dagger) + (D_{\mu}\phi)^\dagger(D^{\mu}\phi) + \frac{1}{2} m^2 \phi^\dagger\phi - \lambda (\phi^\dagger\phi)^2 \label{eq_higgs}
\end{eqnarray}
where $G$, $W$, and $B$ are the gauge field tensors, $D_{\mu}$ are the corresponding gauge covariant derivatives, $q^L$ and $q^R$ are the left- and right-handed quark fields of six flavors, and $L_j$ are the left-handed $SU(2)_L$ doublets for three generations of both quarks and leptons while $R_j$ are the right-handed singlets. In the Higgs part of the Lagrangian, $\phi$ is the Higgs field, $m$ is the Higgs mass, $\lambda$ is a dimensionless constant of about 1/8 in SM, and $y_j$ is the Yukawa coupling for the mass term of the fermion field $\psi_j$, whereas it is still unclear if neutrinos should be included in such Dirac mass terms.
}

The discrete $Z_2$ mirror symmetry relates the SM fields to their mirror partners. Without loss of generality, we assume that left-(right-)handed fermion fields are odd (even) under mirror transformation $\mathcal{M}$ as follows,
\begin{equation}\label{eq_m}
\mathcal{M}: \: \psi_L \rightarrow -\psi'_L, \: \psi_R \rightarrow \psi'_R, \: \phi \rightarrow -\phi'
\end{equation}
where the negative sign in the transformation of the Higgs field $\phi$ ensures the Yukawa mass terms invariant under mirror transformation, which is obvious for the composite Higgs as shown in the next section. All the other fields are transformed trivially under $\mathcal{M}$. Note that mirror transformation is analogous to the chirality operator $\gamma^5$ but connecting the two sectors instead of working within one. Furthermore, the mirror symmetry is closely related to the axial symmetry and the mirror symmetry breaking at the electroweak phase transition could also trigger a series of axial symmetry breakdowns of $U_A(1)$ and $SU_A(2)$ \cite{tan2019c} that will be discussed in the next section.

The total Lagrangian of SM$^3$ can then be written as
\begin{equation}
\mathcal{L} = \mathcal{L}_{\text{SM}}(G,W,B,\psi_L,\psi_R,\phi) + \mathcal{L}'_{\text{SM}}(G',W',B',-\psi'_L,\psi'_R,-\phi')
\end{equation}
which is formally symmetric under mirror transformation $\mathcal{M}$. In previous studies on mirror symmetry, $\mathcal{M}$ has often been confused with $CP$- or $P$-like transformations without recognizing its chiral oddity as shown in Eq. (\ref{eq_m}) \cite{berezhiani2004,foot2014,chacko2006}.
Mirror transformation $\mathcal{M}$ can be combined with $CP$ operation to relate ordinary fermions to mirror antifermions as follows,
\begin{equation}
\mathcal{M}CP: \: \psi_L \rightarrow -i\gamma^2\bar{\psi}^{'tr}_R, \: \psi_R \rightarrow i\gamma^2\bar{\psi}^{'tr}_L, \: \phi \rightarrow -\phi',
\end{equation}
and it can also be used with parity transformation as,
\begin{equation}
\mathcal{M}P: \: \psi_L \rightarrow -\gamma_0\psi'_R, \: \psi_R \rightarrow \gamma_0\psi'_L, \: \phi \rightarrow -\phi^{'\dagger}
\end{equation}
where again the left- and right-handed fermion fields are transformed differently (by a negative sign). The total Lagrangian is formally invariant as well under both $\mathcal{M}CP$ and $\mathcal{M}P$ transformations.

The spontaneous mirror symmetry breaking (SMSB) degenerates the weak interaction into $SU_L(2)$ and $SU'_R(2)$ causing maximal breaking in parity and mirror symmetries in the weak interaction. It also results in slightly uneven vacuum expectation values of Higgs fields in the two sectors \cite{tan2019} that provide the very small violations in transformations like $\mathcal{M}P$ and $CP$. The strong and electromagnetic forces do not distinguish left-handed fields from right-handed ones. Furthermore, masses are generated via quark condensation and the strong interaction as discussed in the next section. Therefore, the interaction basis and the mass basis are aligned and there is no cross-sector particle mixing in these interactions. Notwithstanding, the weak interaction of $SU_L(2)$ and $SU'_R(2)$, affected by SMSB, can mix particles between the two sectors. For example, left-handed fermion $\psi_L$ could be mixed with $\psi'_L$ under $SU_L(2)$ and similarly $\psi'_R$ mixed with $\psi_R$ under $SU'_R(2)$. However, due to the neutrino degeneracy as discussed below, all left-handed neutrinos are in the ordinary sector and all right-handed ones are in the mirror sector. The result is that neutrinos can not be mixed between the two sectors and thus no cross-sector mixing is possible for charged leptons either.

The only possible cross-sector mixing in the weak interaction naturally occurs between quarks and mirror quarks. This results in non-vanishing matrix elements of $V_{qq'} \sim 0.1$ in the extended unitary CKM matrix as follows \cite{tan2019d},
\begin{equation} \label{eq_qmix}
V_{qmix} = 
\begin{pmatrix}
V_{ud} & V_{us} & V_{ub} & V_{uu'} \\
V_{cd} & V_{cs} & V_{cb} & V_{cc'} \\
V_{td} & V_{ts} & V_{tb} & V_{tt'} \\
V_{dd'} & V_{ss'} & V_{bb'} & V' \\
\end{pmatrix}
\end{equation}
where, for simplicity, mirror part $V'$ that is analogous to the $3\times3$ SM CKM matrix is not expanded explicitly and other cross-sector elements ($V_{ij'}$ for $i\neq j$) are assumed to vanish (at least in the first order) and hence suppressed. Note that Eq. (\ref{eq_qmix}) is introduced formally in consideration of unitarity as the original $3\times 3$ CKM matrix is not unitary under the new model \cite{tan2019d}. Both symmetry breaking phases of $CP$-violation and SMSB in the extended CKM matrix are a result of SMSB and therefore their symmetry breaking scales should be comparable. Indeed, such an intuitive estimate is consistent with experimental evidence \cite{tan2019c} and helpful for calculating vacuum energy below. Due to this type of quark mixing, the amazing messenger channels between the two sectors via neutral particle oscillations such as $n-n'$ \cite{tan2019} and $K^0-K^{0'}$ \cite{tan2019c} are possible leading to many important results and applications \cite{tan2019,tan2019a,tan2019b,tan2019c,tan2019d}. Note that this mirror mixing mechanism is similar to that of the generation mixing for quarks and neutrinos. However, it is evident that quarks and mirror quarks are charged differently and therefore the $qq'$ mixing does not manifest at the single-quark level because of charge conservation. The neutral hadron oscillations could then be understood as a type of time-like topological transitions in contrast to other known nonperturbative transitions with space-like barriers such as instantons, sphalerons, and quarkitons \cite{tan2019c}.

Another consequence of SMSB is the degeneracy of neutrinos, i.e., the two sectors share the same three generations of neutrinos. More specifically, $\nu_R$ (singlet in the ordinary sector) is the same as $\nu'_R$ ($SU'_R(2)$ doublet in the mirror sector) and vice versa for left-handed neutrinos ($\nu_L = -\nu'_L$). In other words, right-handed neutrinos do not participate in any gauge interactions in the ordinary sector and neither do left-handed neutrinos in the mirror sector. As further explored in the next section, SMSB gives the neutrinos tiny but non-vanishing masses. Under the supersymmetry principle that will be discussed later, the neutrinos are Dirac fermions instead of Majorana type.

Note that SM$^3$ is different from other mirror-matter models \cite{berezhiani2004,foot2014}. There is no explicit interaction between ordinary and mirror matter particles in SM$^3$. Secondly, left- and right-handed fermions are transformed oppositely under the mirror symmetry and it is this chiral oddity of the mirror symmetry that results in a nearly perfect cancellation in vacuum expectation values of the scalar fields and consequently very tiny vacuum energy density. Thirdly, unlike other models, the mechanism of neutral particle-mirror particle oscillations in SM$^3$ is due to the cross-sector quark mixing instead of some arbitrarily introduced interaction. Last but not least, SM$^3$ can naturally incorporate the four-fermion interactions for staged quark condensation or phase transitions including the top quark condensation to trigger both electroweak and mirror symmetry breakdowns. Most remarkably, this rather exact mirror-matter model can naturally explain a wide variety of celebrated puzzles in physics \cite{tan2019,tan2019a,tan2019b,tan2019c,tan2019d}. SM$^3$ can be tested and its parameters (mixing angles and the mass splitting scale) can be better measured in various experiments proposed in Ref. \cite{tan2019d}.

Our earlier studies have revealed that the mirror symmetry breaking scale is $\delta v/v = \delta m/m \sim 10^{-15} \text{--} 10^{-14}$ \cite{tan2019,tan2019a,tan2019b,tan2019c,tan2019d}. This tiny breaking scale is required in M$^3$ for consistent explanation of various scenarios, e.g., neutron lifetime anomaly and dark-to-baryon matter ratio \cite{tan2019}, evolution of stars \cite{tan2019a}, ultra-high energy cosmic rays \cite{tan2019b}, and matter-antimatter imbalance \cite{tan2019c}. For example, experimental evidence from neutron lifetime measurements and observed dark-to-baryon ratio of 5.4 and baryon-to-photon number density ratio of $n_B/n_\gamma = 6.1\times10^{-10}$ constrain the $n-n'$ mass splitting scale to be about $10^{-6}$ -- $10^{-5}$ eV \cite{tan2019,tan2019c} corresponding to a relative scale of $\delta m/m \sim 10^{-15} \text{--} 10^{-14}$. 

Considering that hadrons (quark condensates) acquire their masses similarly due to the strong interaction and SMSB, the relative mass splitting scale should be universal at least for all hadrons and Higgs VEVs. As indicated in Ref. \cite{tan2019c} and discussed above, the $CP$ violation is also originated from the spontaneous mirror symmetry breaking. From the measured mass difference of  $\Delta_{K^0_SK^{0}_L} = 3.5 \times 10^{-6}$ eV in $CP$-violating $K^0$ oscillations, we can obtain the symmetry breaking scale of $\delta m/m \sim 10^{-14}$ from the well known kaon mass $m(K^0) = 497.6$ MeV. Further studies assuming equivalence of the $CP$ violation and mirror symmetry breaking scales have been done for precise predictions of invisible decays of long-lived neutral hadrons \cite{tan2020d}.

Eq. (\ref{eq_rvac}) provides a good estimate of vacuum energy for a single scalar field under quantum field theory as used in the literature \cite{kolb1990}. However, a full description of vacuum energy or the cosmological constant for multiple scalar fields could not be obtained without a valid theory of quantum gravity. Under SM$^3$, the Higgs field $\phi$ and the mirror Higgs $\phi'$, or for that matter the two sectors, are decoupled in terms of gauge interactions leading to separate quartic Higgs terms in the Lagrangian. But they should participate in gravity coherently in the same way. Therefore, it is natural to apply the quantum superposition principle to obtain vacuum energy density as $\langle \phi-\phi' \rangle^4 \sim \delta v^4$ instead of an incoherent sum like $\langle \phi \rangle^4 + \langle \phi' \rangle^4$. In general, the extension of vacuum energy defined by multiple scalar fields can then be expressed as,
\begin{equation}
\rho_{\text{vac}} \sim \langle \sum_i \phi_i \rangle^4 \sim (\sum_i{v_i})^4.
\end{equation}

As shown in Eq. (\ref{eq_m}), the mirror symmetry requires that the two VEVs of the two sectors have opposite signs. Using the fairly well constrained mirror symmetry breaking scale of $10^{-14}$ and the Higgs VEV of $v = 246$ GeV, therefore, we can estimate the vacuum energy density as,
\begin{equation}
\rho_{\text{vac}} \sim (\delta v)^4 \sim (10^{-3}\, \text{eV})^4
\end{equation}
which agrees remarkably well with observation. Further consideration of the spontaneous symmetry breaking as staged quark condensation will be presented in the next section and the effects on vacuum energy will be discussed.

\section{Quark condensation with four-fermion interactions and lepton masses\label{sec_4f}}

The chiral oddity of Higgs under mirror transformation is critical for explanation of dark energy in the previous section. In this section, we shall explore the nature of the Higgs mechanism that can be better understood as the phenomenon of quark condensation. Using the formalism of four-fermion interactions for quark condensation, the simple relations between Higgs mass, VEV, and top quark mass can be astonishingly established. Furthermore, masses of charged leptons and properties of neutrinos are naturally explained. Even deeper underlying principles and conjectures will be discussed in the next section.

The Nambu-Jona-Lasinio mechanism using a four-fermion interaction was applied to study bound quark states in analogy with superconductivity \cite{nambu1961}. The Higgs mechanism was then proved to be just an effective low energy theory due to quark condensation of the four-fermion interaction \cite{eguchi1976,hasenfratz1991}. The idea of using top-quark condensation for the electroweak phase transition was proposed \cite{nambu1988,nambu1988a,miransky1989} and a detailed study to formulate the dynamical symmetry breaking of the standard model with the idea was then carried out \cite{bardeen1990}, indicating that Higgs may be composite instead of a fundamental particle.

Many successful descriptions for meson mass relations \cite{weinberg1995} and effectiveness of the $\sigma$ model \cite{pelaez2016} show strong support for the idea of staged quark condensation leading to SMSB and other symmetry breaking transitions including the electroweak phase transition \cite{tan2019c}. Here we shall present how the four-fermion interactions and staged quark condensation under SM$^3$ lead to the effective Higgs mechanism and new understanding for masses of quarks and leptons.

First, we can introduce a four-quark interaction for a simple case of one quark (e.g., top quark) condensation,
\begin{equation}\label{eq_v4q}
V_{4q} = \frac{2y^2}{m^2} \bar{q}_L q_R\bar{q}_R q_L
\end{equation}
which leads to the Higgs mechanism as shown in Eq. (\ref{eq_higgs}) after condensation with $y$ as the Yukawa coupling and $m$ as the Higgs mass. Such a duality of the four-fermion interaction and the Higgs mechanism has been well studied before \cite{eguchi1976,hasenfratz1991}. Here we provide new insights for the top quark condensation under SM$^3$. Considering the composite Higgs field as a top quark condensate,
\begin{equation}
\phi=\frac{2y}{m^2}\langle \bar{q}_R q_L \rangle, \: \phi^\dagger = \frac{2y}{m^2}\langle \bar{q}_L q_R \rangle,
\end{equation}
it is easy to obtain the Yukawa mass term of $y(\bar{q}_L q_R \phi + h.c.)$ from $V_{4q}$ in Eq. (\ref{eq_v4q}) by contracting one quark pair and the Higgs mass term by condensing both pairs. Another renormalizable term can be constructed by condensing four pairs of quarks in the second order term of $V_{4q}$,
\begin{equation}
\frac{1}{2m^4} \langle V_{4q}^2 \rangle = \frac{1}{8} (\phi^\dagger\phi)^2
\end{equation}
which provides the quartic term in $\mathcal{L}_{\text{Higgs}}$. Note that $y=g=1$ (strong coupling), $m$, and $\lambda = 1/8$ are all bare parameters. The slight mirror symmetry breaking of $\delta v/v \sim 10^{-14}$ makes the quadratic divergence mostly canceled out as $\delta m^2 \propto m^2 - m^{'2} \simeq 10^{-14} m^2$. Therefore, the Higgs and top quark parameters under the protection of the mirror symmetry do not see much radiative corrections. This is well supported by measurements and simple relations of $v^2= 2m^2_t = 4m^2$. In particular, based on well known $v=246.2$ GeV, SM$^3$ predicts that the top quark mass $m_t = 174.1$ GeV and the Higgs mass $m=123.1$ GeV, which are miraculously close to the measured values of $m_t = 173.0$ GeV and $m=125.1$ GeV \cite{particledatagroup2018}. The small variation from the bare parameters could possibly stem from the mixing of top quark with other quarks.

In general, four-fermion interactions involving quark condensates can be written as,
\begin{equation}
V_{4f} = \frac{g_{\psi}g}{m^2_{\bar{q}q}} (\bar{\psi}_L \psi_R\bar{q}_R q_L + \bar{\psi}_R \psi_L\bar{q}_L q_R)
\end{equation}
where $\psi$ is another quark or lepton field, $g_{\psi}$ denotes the coupling between $\psi_L$ and $\psi_R$, and $m_{\bar{q}q}$ is the condensation energy scale (i.e., mass of the Higgs-like particle from $\bar{q}q$ condensation). Here we consider only four-fermion terms that are composed of quarks and leptons of the same flavor. At least one pair of left- and right-handed fermions have to be quarks for condensation. The possible cross-flavor contributions due to higher order corrections are ignored here. 
Considering the case of b-quark and $\tau$-lepton in the same generation and $SU_L(2)$ representation, they both couple to the $\bar{b}b$ condensate in the same way. The only difference is that $\tau$-lepton coupling (between $\tau_R$ and $\tau_L$) from the electromagnetic force gives $g_\tau = \sqrt{4\pi\alpha} \simeq 0.3$. Therefore, we can obtain a mass relation of $m_\tau = 0.3 m_b$ which agrees well with experimental values of $m_\tau = 1.77$ GeV and $m_b = 4.18$ GeV. Similar relations hold also for $e$- and $\mu$-leptons in other generations. The conclusion is that charged leptons have similar masses to their corresponding down-type quarks.

If neutrinos were charged or not degenerate, they would have been as massive as the up-type quarks. After the mirror symmetry breaking, the Yukawa mass term for a neutrino can be obtained using the neutrino degeneracy relations of $\nu_L = -\nu'_L$ and $\nu_R = \nu'_R$,
\begin{equation}
-y(\bar{\nu}_L \nu_R \phi + \bar{\nu}'_L \nu'_R \phi' + h.c.) = -y(\bar{\nu}_L \nu_R (\phi-\phi') + h.c.)
\end{equation}
where the Yukawa coupling $y = g_W = \sqrt{4\pi\alpha_W} \sim 0.65$. Therefore the neutrino mass is obtained as,
\begin{equation}
m_\nu = y \frac{\delta v}{\sqrt{2}} \sim \delta v.
\end{equation}
Pertinent to the top quark condensation, the bare mass of $\tau$-neutrino is $m(\nu_\tau) \sim \delta v_t \sim 10^{-3}$ eV (for simplicity, no distinction is made between flavor and mass eigenstates for neutrinos). Similarly, we can obtain $m(\nu_\mu) \sim \delta v_c \sim 10^{-5}$ eV and $m(\nu_e) \sim \delta v_u \sim 10^{-6}$ eV for $\mu$- and $e$-neutrinos corresponding to c-quark condensation and QCD phase transition (discussed below), respectively. From the squared mass differences of three generation neutrinos measured in oscillation experiments \cite{particledatagroup2018}, we can infer that two of them should have masses of about $10^{-2}$ eV and $10^{-3}$ eV while the lightest one could even be massless. Our results indicate that even the lightest neutrino must have mass (possibly $\sim 10^{-4}$ eV considering radiative corrections). It is impressive that our bare mass estimates are within one or two orders of magnitude of these so tiny masses. Unlike the top quark mass, radiative corrections for these tiny neutrino masses should be much more significant, which may explain the discrepancies.

As suggested in Ref. \cite{tan2019c}, the phase transitions due to quark condensation may be staged. This can be understood under the framework of four-fermion interactions assuming different mass scales for condensation of different quark flavors. Imagining an initial $U(6)$ flavor symmetry for fermions, we can break it down following a series of chiral symmetry breaking processes \cite{tan2019c},
\begin{eqnarray}
U(1) \times SU(6) &/& U^t_A(1) \rightarrow U(1) \times SU(5) \times  U^t_V(1) \nonumber \\
&/& U^b_A(1) \rightarrow U(1) \times SU(4) \times  \prod_{i=t,b}U^i_V(1) \nonumber \\
&/& U^c_A(1) \rightarrow U(1) \times SU(3) \times  \prod_{i=t,b,c} U^i_V(1) \nonumber \\
&/& U^s_A(1) \rightarrow U(1) \times SU(2) \times  \prod_{i=t,b,c,s} U^i_V(1) \nonumber \\
&/& U_A(2) \rightarrow U_V(1) \times SU_V(2) \times \prod_{i=t,b,c,s} U^i_V(1)
\label{eq_condense}
\end{eqnarray}
where subgroup properties of $SU(N) \supset SU(N-1) \times U(1)$, $SU(N) = SU_L(N) \times SU_R(N) = SU_V(N) \times SU_A(N)$, and $U(1) = U_V(1) \times U_A(1)$ are applied. Among the final preserved symmetries, $SU_V(2)$ is the isospin symmetry, $U_V(1)$ is for baryon conservation, and the other vector $U(1)$ symmetries are for conservation of top, bottom, charm, and strange numbers in the strong interaction, respectively. Evidently, these symmetries are eventually broken as well due to the electroweak interaction. Further details on this new understanding of staged phase transitions, associated topological ``quarkiton'' processes, and their impact on baryogenesis in the early universe, can be found in Ref. \cite{tan2019c}.

According to the hierarchy of quark masses, the energy scales (VEVs of Higgs-like fields) are estimated to be on the order of $10^{11}$ eV, $10^{10}$ eV, $10^{9}$ eV for top, bottom, charm condensation stages, respectively, and about $10^{8}$ eV for the last two stages of condensation (QCD phase transition) in Eq. (\ref{eq_condense}) \cite{tan2019c}. At each stage, the condensed quarks provide a Higgs-like composite scalar that can further modify the vacuum structure. Assuming a universal mirror symmetry breaking scale of $\delta v/v \sim 10^{-14}$, we can obtain the net contribution of each stage to the vacuum energy as $\delta v_t \sim 10^{-3}$ eV, $\delta v_b \sim 10^{-4}$ eV, $\delta v_c \sim 10^{-5}$ eV, and $\delta v_s \simeq \delta v_{ud} \sim 10^{-6}$ eV. Under this scenario, the dark energy density can be estimated as,
\begin{equation}
\rho_{\text{vac}} \sim (\sum_i \delta v_i)^4 \sim (10^{-3}\, \text{eV})^4
\end{equation}
which is clearly dominated by the top quark condensation. The new understanding of phase transitions does not change the conclusion on dark energy from the simple picture presented in the previous section.

\section{Maximally interacting, supersymmetric, and mirrored world\label{maxsusy}}

The Feynman path integral formalism can be used to shed light on the principle of a maximally interacting world. The main idea of the path integral approach is that the amplitude of any physical quantity is given by a coherent sum of all possible paths weighted by a phase factor of $\exp(i\mathcal{S}/\hbar)$ where $\mathcal{S}$ is the action as defined by the integration of the Lagrangian over four-dimensional spacetime of our world,
\begin{equation}
\mathcal{S} = \int d^4x \mathcal{L}.
\end{equation}
Now we can formulate the conjecture of a maximally interacting world, which hypothesizes that the action or Lagrangian of a path allows all symmetry-obeying dimension-4 or higher terms. However, for any path with higher order terms (say, dimension-5 or higher and non-renormalizable) in its Lagrangian, its contribution to the amplitude will essentially vanish due to the divergence in the phase factor. Therefore, only paths with renormalizable dimension-4 terms in the Lagrangian can contribute. And of course, these terms also have to satisfy the requirements of gauge and other symmetries. Higher order terms can only manifest when its dimensionality is reduced, e.g., by condensing higher-dimension fields (fermions) into lower-dimension ones (bosons).

For example, the dimension-6 four-fermion terms emerge only after the phase transition of quark condensation. It turns into the fermion and Higgs mass terms when its dimensionality is reduced to four by quark condensation. Similarly, the Higgs kinetic term can be obtained from condensation of the dimension-8 operator of $\propto(D^{\mu}(\bar{q}_R q_L))^\dagger D_{\mu}(\bar{q}_R q_L)/m^4$. Even the second order of a four-fermion term takes into effect by condensing into the quartic Higgs term (from dimension-12 to dimension-4). Quark condensation with four-fermion interactions becomes a natural mechanism for SMSB under the hypothesis of a maximally interacting world.

Supersymmetry (SUSY) is a desired feature in ultimate unification theories such as string theory. However, a new understanding of SUSY may be needed. Nambu proposed a quasi-SUSY principle when he observed the matching of degrees of freedom (DoF) between fermions and bosons in many models \cite{nambu1988,nambu1988a}. Instead of counting DoF for one generation of fermions as Nambu did, more detailed and realistic counts of DoF for SM$^3$ are presented in Table \ref{tab_dof}. Results for one sector are presented as DoF counts in both ordinary and mirror sectors are identical. After SMSB,  degeneracy of neutrinos causes their DoF reduced by half assuming they are Dirac particles. Meanwhile, $W^\pm$ and $Z^0$ bosons acquire masses and consequently one extra degree of freedom each. The global $U(6)$ flavor symmetry breaking as shown in Eq. (\ref{eq_condense}) produces 63 pseudo-Nambu-Goldstone bosons (pNGB). Here the leftover symmetries of vector $SU(2)$ and $U(1)$ preserve eight degrees of freedom while the broken $U_A(1)$ is dynamically canceled by other flavor $U_A(1)$'s \cite{tan2019c} and therefore associated with no boson. After a detailed counting as shown in Table \ref{tab_dof}, a pseudo-SUSY of $n_f=n_b=90$ in each sector is observed and it is called ``pseudo'' because the counted pNGB and Higgs particles are massive, composite, and not fundamental.

Before SMSB, all particles are fundamental and massless. The flavor $U(6)$ has to be restored as a gauge symmetry under some unknown mechanism for more gauge bosons. The SUSY principle of $n_f=n_b=96$ in each sector is then obeyed as shown in Table \ref{tab_dof}. The unbroken SUSY can thus be understood as the DoF symmetry between fundamental fermions and associated gauge bosons. Instead of SUSY, it is the mirror symmetry that makes a mirrored copy of ordinary particles and, moreover, it is SMSB that simultaneously broke SUSY to a pseudo-SUSY as discussed above.

\begin{table*}
\caption{\label{tab_dof}Numbers of degrees of freedom (DoF) for both fermions and bosons in one sector are shown and compared before and after the spontaneous mirror symmetry breaking (SMSB), obeying SUSY ($n_f=n_b=96$) and pseudo-SUSY ($n_f=n_b=90$) symmetries, respectively.}
\begin{ruledtabular}
\begin{tabular}{l | l r | l r}
& fermions & DoF & bosons & DoF \\
\hline
after & quarks & $2\times2\times3\times6 = 72$ & gluons & $2\times8=16$ \\
SMSB & $e$,$\mu$,$\tau$ & $2\times2\times3=12$ & EW gauge bosons & $3\times3+1\times2=11$ \\
& neutrinos & $2\times3 = 6$ & U(6) pNGB & $36\times2-9 = 63$ \\
& total $n_f$ in one sector & 90 & total $n_b$ in one sector & 90 \\
\hline
before & quarks & $2\times2\times3\times6 = 72$ & gluons & $2\times8=16$ \\
SMSB & $e$,$\mu$,$\tau$ & $2\times2\times3=12$ & EW gauge bosons & $3\times2+1\times2=8$ \\
& neutrinos & $2\times2\times3 = 12$ & U(6) gauge bosons & $36\times2 = 72$ \\
& total $n_f$ in one sector & 96 & total $n_b$ in one sector & 96 \\
\end{tabular}
\end{ruledtabular}
\end{table*}

Under the (pseudo)-SUSY principle, neutrinos have to be Dirac fermions instead of Majorana type in order to balance degrees of freedom of fermions and bosons. As discussed earlier, these neutrinos obtain tiny Dirac masses via slightly broken mirror symmetry, which in turn supports such a conjecture. More intriguingly, the SUSY principle may be the reason why such three generations of fermions and such a gauge group of $U_f(6)\times SU_c(3)\times SU_w(2)\times U_Y(1)$ before the symmetry breaking are selected by nature. What other SUSY-obeying choices of fermions and gauge groups are possible in the energy desert between the electroweak scale of $10^2$ GeV and the Planck scale of $10^{19}$ GeV? Further studies on the flavor $U(6)$ and other possible gauge symmetries may reveal more fascinating results under the new SUSY principle.

In SM$^3$, there is no fundamental scalar field. The Higgs-like fields from quark condensation are composite and have odd mirror parity under mirror transformation. A true scalar field may exist only shortly after the Big Bang when it is speculated that the universe goes through an inflationary era, i.e., a period of exponential cosmological expansion to be consistent with observed isotropy and flatness of the universe \cite{kolb1990}. The large VEV (close to the Planck scale) of such a field can indeed provide an enormous amount of vacuum energy and hence drive the inflation, possibly for both ordinary and mirror sectors, respectively \cite{hodges1993}. Then the scalar fields decay into massless fermions and bosons with equal numbers of degrees of freedom following the SUSY, mirror, and gauge symmetries. One interesting aspect is related to the energy scale of about $10^{16}$ GeV (similar to the GUT scale) for typical inflation models. We can assume that the mirror symmetry breaking scale starts with an order of one in the beginning of the inflation. After the inflation and at the electroweak phase transition of about $10^2$ GeV,  the energy scale is changed by 14 orders of magnitude and the mass splitting scale for mirror symmetry could also be reduced by the same factor. This could explain why a tiny mass breaking scale of $\delta v/v \sim 10^{-14}$ is observed today. Another intriguing consequence is that the universe could have a double inflation process by both ordinary and mirror scalar fields separately, which could result in a dipole component in the acceleration of cosmic expansion. This might explain the recent discovery of a large dipole component of cosmic acceleration in a reanalysis of type Ia supernova data by Colin \textit{et al.} \cite{colin2019}.

Such scenarios are particularly supported considering that the mirror matter temperature $T'$ may be different from the ordinary matter temperature $T$ with a likely ratio of $T'/T \sim 1/3$ as indicated by various studies \cite{kolb1985,hodges1993,berezhiani2004,foot2014,tan2019,tan2019b}. When SMSB occurs slightly earlier in the mirror sector, the net VEV of the vacuum can be as large as $10^2$ GeV for a brief period of time. This can cause a mini-inflation, though negligible compared to the early one, which may have effects on subsequent baryogenesis and nucleosynthesis. Even smaller-scale mini-inflation processes due to staged quark condensation may contribute later as well. More intriguingly, such mini-inflation processes or similar condensation mechanism later may be the source of the so-called Hubble tension, i.e., the $>4\sigma$ discrepancy of the Hubble constant between the CMB data \cite{planckcollaboration2018} and the local measurements \cite{riess2019}. Replacing $\Lambda$CDM with new understanding of $\Lambda$ and self-interacting mirror matter as in SM$^3$, the new standard cosmology model ($\Lambda$M$^3$) could help us better understand the early universe. Most recent development of the new model into a hierarchy of supersymmetric mirror models has been done for explaining the dynamic evolution of the Universe \cite{tan2020,tan2020a} and understanding the nature of black holes \cite{tan2020b}.

Imagining a maximally interacting, supersymmetric, and mirrored world after the inflation, the evolution of the universe is governed by $\Lambda$M$^3$ and SM$^3$ with slightly broken mirror symmetry, which can strikingly and consistently address many big questions on topics such as dark energy, dark matter, baryogenesis, evolution of stars, neutron lifetime anomaly, etc \cite{tan2019,tan2019a,tan2019b,tan2019c,tan2019d}. A perfectly imperfect mirror symmetry is indeed the key to unlock the beauty and elegance of our universe.

Given the hierarchy of quark masses, SM$^3$ can provide a natural explanation of lepton masses including tiny neutrino masses. Nonetheless, mass hierarchies between quark flavors are as yet to be understood. Why is the $U(6)$ flavor symmetry broken in stages instead of all at once so that all quarks acquire the same mass? How are such mass scales embedded in $U(6)$ that was gauged before SMSB? What is the role of the strong $SU_c(3)$ interaction in the quark mass hierarchy?
Many such questions still await future investigation.

\begin{acknowledgments}
This work is supported in part by the National Science Foundation under
grant No. PHY-1713857 and the Joint Institute for Nuclear Astrophysics (JINA-CEE, www.jinaweb.org), NSF-PFC under grant No. PHY-1430152.
\end{acknowledgments}

\bibliography{vac}

\end{document}